\documentclass[twocolumn]{aastex62}

\usepackage{graphicx} 
\usepackage{float} 
\usepackage{natbib}
\usepackage{amsmath}
\usepackage[T1]{fontenc}
\usepackage[utf8]{inputenc}

\shorttitle{Origin of Kepler-419b}
\shortauthors{Jackson, Dawson, \& Zalesky}

\begin{document}
\title{The Origin of Kepler-419b: A Path to Tidal Migration Via Four-body Secular Interactions}

\correspondingauthor{Jonathan M. Jackson}
\email{jqj5401@psu.edu}

\author{Jonathan M. Jackson}
\affil{Department of Astronomy \& Astrophysics, Center for Exoplanets and Habitable Worlds, The Pennsylvania State University, University Park, PA 16802}
\author{Rebekah I. Dawson}
\affil{Department of Astronomy \& Astrophysics, Center for Exoplanets and Habitable Worlds, The Pennsylvania State University, University Park, PA 16802}
\author{Joseph Zalesky}
\affil{School of Earth \& Space Exploration, Arizona State University, Tempe, AZ 85287}

\begin{abstract}
We test the high-eccentricity tidal migration scenario for Kepler-419b, a member of the eccentric warm Jupiter class of planets  whose origin is debated. Kepler-419 currently hosts two known planets (b,c). However, in its current configuration, planet c cannot excite the eccentricity of planet b enough to undergo high-eccentricity tidal migration. We investigate whether the presence of an undiscovered fourth body could explain the orbit of Kepler-419b. We explore the parameter space of this potential third giant planet using a suite of N-body simulations with a range of initial conditions. From the results of these simulations, coupled with observational constraints, we can rule out this mechanism for much of the parameter space of initial object d conditions. However, for a small range of parameters (masses between 0.5 and 7 $m_{\rm{Jup}}$, semi-major axes between 4 and 7.5 AU, eccentricities between 0.18 and 0.35, and mutual inclinations near 0$^{\circ}$) an undiscovered object d could periodically excite the eccentricity of Kepler-419b without destabilizing the system over 1 Gyr while producing currently undetectable radial velocity and transit timing variation signals.
\end{abstract}



\section{Introduction} \label{intro}

\begin{figure}
\center{\includegraphics{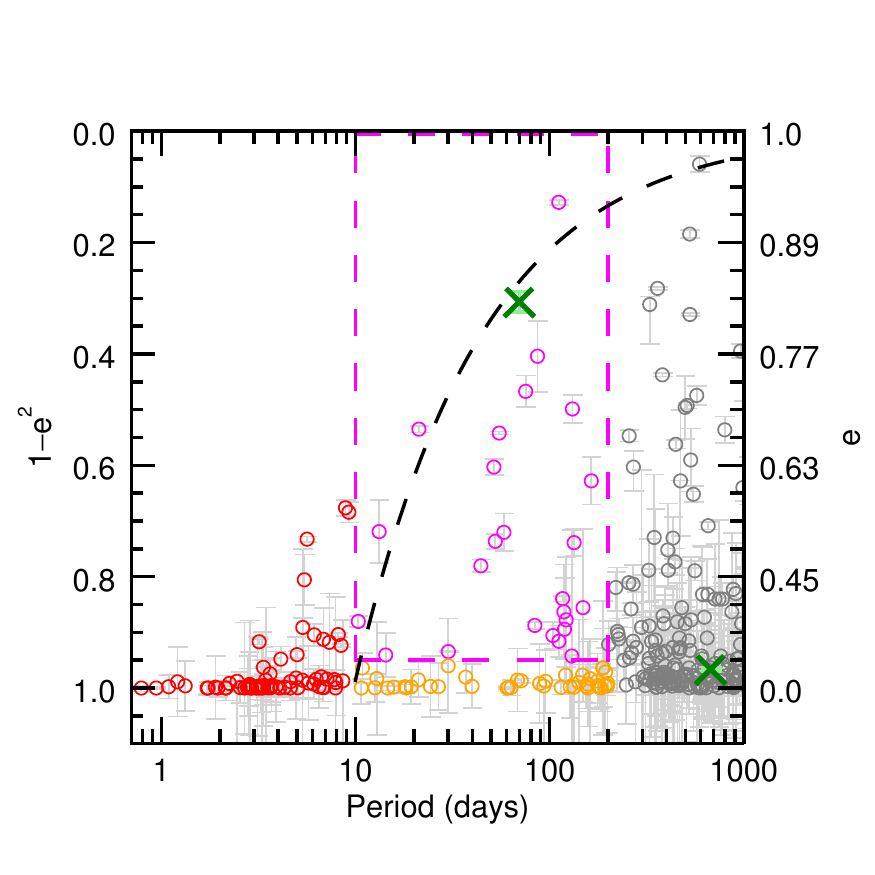}}
\caption{Kepler-419b is one of a population of Jupiter-mass planets with intermediate periods and large eccentricities. Each point represents a currently known exoplanet with $m>0.5M_{Jup}$, separated into cold Jupiters (gray), low-eccentricity warm Jupiters (orange), hot Jupiters (red), and, of interest to this paper, eccentric warm Jupiters (purple), which have periods between 10 and 200 days, and eccentricities greater than 0.2. Kepler-419b and Kepler-419c are denoted by green Xs. The dashed black line represents the tidal circularization track of constant angular momentum that would produce a hot Jupiter with a period of 10 days. If Kepler-419b were to migrate via high-eccentricity migration, its eccentricity would need to at least surpass that line. Data taken from exoplanets.org \citep{Han14}.}
\label{fig:ecc}
\end{figure}

Both hot and warm Jupiters are believed to form beyond the ice-line and migrate inward to their current semi-major axes (\citealt{Bod00,Raf06}; see \citealt{Daw18} for a review of hot Jupiter origins theories). Although disk migration \citep{Gol80,War97,Bar14} could deliver hot and warm Jupiters, it is difficult to reconcile disk migration with warm Jupiters' eccentricity distribution (Figure \ref{fig:ecc}) because planet-disk interactions tend to damp eccentricities \citep{Bit13,Dun13}. Planet-disk interactions can sometimes excite eccentricities, but they typically saturate at a random velocity equal to the sound speed ($e\lesssim 0.03$ for a 100 day orbit; \citealt{Duf15}). Eccentricity growth could occur after the disk migration stage via scattering, but such growth is limited by $v_{\rm{escape}}/v_{\rm{keplerian}}$, making it difficult for planets on close in orbits to attain large eccentricities \citep{Gol04,Ida13,Pet14}. The disconnect between formation theories and observations makes warm Jupiters an interesting parameter space for testing migration mechanisms.

Eccentric warm Jupiters may have arrived through high-eccentricity tidal migration (e.g., \citealt{Hut81,Wu03}). \citet{Pet16} concluded that high-eccentricity tidal migration triggered by an outer planetary companion can account for $\sim20\%$ of warm Jupiters and most warm Jupiters with $e\geq0.4$. In this mechanism, planets are excited to large orbital eccentricities such that tidal friction with their host star works to circularize the planet's orbit and reduce the semi-major axis. The planet's angular momentum remains constant once it is decoupled from other perturbing bodies. Via high-eccentricity migration, many warm Jupiters would end their lives as hot Jupiters on circular orbits. Several theories have been proposed to explain the original excitation of eccentricity, including planet-planet scattering \citep{Ras96}, secular chaos \citep{Wu11}, and stellar flybys (e.g., \citealt{Kai13}). These mechanisms are successful in producing a wide distribution of giant planet eccentricities from multi-planet systems with initially circular orbits (e.g., \citealt{Jur08,Cha08}). 

Here we focus on secular interactions between planets and their ability to induce high-eccentricity tidal migration. While some planets, such as HD 80606 b \citep{Mou09} and HD 17156 b \citep{Bar09}, have very large eccentricities and may currently be undergoing tidal migration en route to becoming a hot Jupiter, the intermediate eccentricity warm Jupiters would all need smaller angular momenta (or, equivalently, larger eccentricities at a given semi-major axis) than we observe to be undergoing tidal migration. One effect that could induce this migration is for a perturbing body to periodically excite the planet's eccentricity through secular interactions (e.g., \citealt{Don14}) leading to migration by tidal friction during the high-eccentricity phase. We would be observing the planet during the low eccentricity portion of its eccentricity oscillation cycle. There is some evidence for this mechanism in the trend that eccentric warm Jupiters are more likely to have companions \citep{Don14,Bry16} than circular warm Jupiters. The Kozai-Lidov mechanism \citep{Wu03,Nao11} is an extreme case of this type of secular eccentricity modulation, but typically requires large mutual inclinations (see \citealt{Li14} for the case of an eccentric, coplanar Kozai perturber).

We focus on an interesting system for testing theories of tidal migration, Kepler-419. Kepler-419 is a $1.39^{+0.08}_{-0.07}M_{\sun}$ star with two known planetary companions, Kepler-419b and Kepler-419c. Kepler-419b is a transiting warm Jupiter at semi-major axis $a=0.370^{+0.007}_{-0.006}$ AU with a large eccentricity of $e=0.833\pm0.013$ measured via the ``photoeccentric effect" \citep{Daw12J} and confirmed via radial velocity (RV) measurements \citep{Daw14a,Alm18}. Kepler-419c is a $m=7.3\pm0.4M_{Jup}$ perturbing body with moderate eccentricity ($e=0.184\pm0.002$) located at $a=1.68\pm0.03$ AU.

\citet{Daw14a} used RV measurements from Keck HIRES to measure the mass of planet b ($m=2.5\pm0.3M_{Jup}$) and improve precision on the eccentricity value. They used transit timing variations (TTVs) of planet b to deduce precise physical and orbital parameters of planet c. The linear trend in the RV curve is consistent with the properties of planet c derived from the TTVs. Limits on transit duration variations provided no additional constraints on planet c's properties. The full set of system parameters calculated by simultaneously fitting the RVs and TTVs are presented in Table \ref{tab:kep}, an adaptation of Table 4 in \citet{Daw14a}. \citet{Alm18} used RV measurements from SOPHIE \citep{Per08,Bou09} to independently confirm the presence of Kepler-419c. \citet{Saa18} uniquely constrained the masses of both planets using the TTV data alone.

\begin{deluxetable}{ll}
\tablewidth{3in}
\tablecaption{Planet Parameters for Kepler-419b and Kepler-419c at Epoch BJD 2455809.4009671761741629 \citep{Daw14a} \label{tab:kep}}
\tablehead{
\colhead{Parameter} & \colhead{Value}}
\startdata
Stellar mass, $m_{\star}\left(m_{\sun}\right)$ & $1.39^{+0.08}_{-0.07}$ \\
Stellar radius, $R_{\star}\left(R_{\sun}\right)$ & $1.75^{+0.08}_{-0.07}$ \\
Mass b, $m_b\left(m_{\rm{Jup}}\right)$ & $2.5\pm0.3$ \\
Period b, $P_b\left(\rm{days}\right)$ & $69.7546^{+0.0007}_{-0.0009}$ \\
Semi-major axis b, $a_b\left(\rm{AU}\right)$ & $0.370^{+0.007}_{-0.006}$ \\
Eccentricity b, $e_b$ & $0.833\pm0.013$ \\
Argument of pericenter b, $\omega_b\left(^{\circ}\right)$ & $95.2^{+1.0}_{-1.2}$ \\
Mean anomaly b, $M_b\left(^{\circ}\right)$ & $68.69\pm0.05$ \\
Inclination b, $i_b\left(^{\circ}\right)$ & $88.95^{+0.14}_{-0.17}$ \\
Longitude of ascending node b, $\Omega_b\left(^{\circ}\right)$ & 0 (fixed) \\
Longitude of pericenter b, $\varpi_b\left(^{\circ}\right)$ & $95.2^{+1.0}_{-1.2}$ \\
Mass c, $m_c\left(m_{\rm{Jup}}\right)$ & $7.3\pm0.4$ \\
Period c, $P_c\left(\rm{days}\right)$ & $675.47\pm0.11$ \\
Semi-major axis c, $a_c\left(\rm{AU}\right)$ & $1.68\pm0.03$ \\
Eccentricity c, $e_c$ & $0.184\pm0.002$ \\
Argument of pericenter c, $\omega_c\left(^{\circ}\right)$ & $275.3^{+1.2}_{-1.0}$ \\
Mean anomaly c, $M_c\left(^{\circ}\right)$ & $345.0\pm0.3$ \\
Inclination c, $i_c\left(^{\circ}\right)$ & $88^{+3}_{-2}$ \\
Longitude of ascending node c, $\Omega_c\left(^{\circ}\right)$ & $4\pm12$ \\
Longitude of pericenter c, $\varpi_c\left(^{\circ}\right)$ & $279\pm{12}$ \\
Mutual inclination, $i_{\rm{mut}}\left(^{\circ}\right)$ & $9^{+8}_{-6}$ \\
$\omega_b-\omega_c\left(^{\circ}\right)$ & $179.8\pm0.6$ \\
$\varpi_b-\varpi_c\left(^{\circ}\right)$ & $176\pm12$ \\
\enddata
\tablecomments{All orbital elements are computed from Jacobian Cartesian coordinates, ordered from the innermost object outward.}
\end{deluxetable}

At least one other system is consistent with the eccentricity modulated tidal migration scenario: Kepler-693, which hosts an eccentric warm Jupiter and, unlike Kepler-419c, a companion on a high mutual inclination orbit that induces large Kozai-Lidov eccentricity oscillations \citep{Mas17}. However, \citet{Daw15} found a paucity of high-eccentricity warm Jupiters compared with that expected if hot Jupiters obtain their low-$a$ orbits through tidal friction with their host star \citep{Soc12}, suggesting high-eccentricity tidal migration might not be a common migration channel.

If high-eccentricity tidal migration can be ruled out for Kepler-419b, it could further support that this mechanism is not ubiquitous among the warm Jupiter population. Kepler-419b's present day eccentricity is too low for tidal migration and \citet{Daw14a} found that Kepler-419c cannot drive up the eccentricity high enough to induce tidal migration. Secular interactions with planet c do cause oscillations in the eccentricity of planet b, but it is currently in the high-$e$ phase of this cycle and never reaches $a(1-e^2)<0.1$ (represented by the black dotted line in Figure \ref{fig:ecc} and the green shaded area in Figure \ref{fig:two}, the time evolution of Kepler-419b in the two-planet solution). However, before ruling out tidal migration, we need to investigate the possibility that an undetected third planet or a brown dwarf in the system could allow for high-eccentricity migration by increasing the maximum value of Kepler-419b's eccentricity. 

We explore the parameter space for an unseen additional perturber in the Kepler-419 system that could excite the eccentricity of planet b enough for it to periodically reach the large values necessary to undergo tidal migration. We start with the two known planets in their present configuration and simulate the system forward in time. In the tidal migration scenario, the eccentricity and semi-major axis of planet b would decrease over time, while the parameters of planet c remain roughly constant. Any instabilities in the four-body system today would have been stronger in the past when the planets were more tightly packed, allowing us to assume that system configurations that go unstable in our simulations would not have been stable in the past. Similarly, if a set of system parameters fails to sufficiently excite the eccentricity of planet b in the present configuration, we can be confident that it did not do so in the past. Strong precession due to general relativity, tides, or rotational quadrupolar bulges would be needed to decouple planet b from secular interactions with the perturber (e.g., \citealt{Wu03}). With Kepler-419b's current orbital configuration, none of these are strong enough to do so \citep{Daw14a}, nor would they have been if Kepler-419b's semi-major axis was larger in the past.

We test the feasibility of high-eccentricity tidal migration through a suite of $N$-body simulations of the Kepler-419 system with various initial conditions for a possible object d. We motivate our analysis and outline our approach in Section \ref{sims}. We present the results of our simulations in Section \ref{results}. In Section \ref{observations}, we assess how object d would affect the observables of the system. We summarize our conclusions, discuss alternative scenarios, and propose future efforts in Section \ref{conclusion}.

\section{Overview of Simulations} \label{sims}

We seek to test the parameter space of a potential object d in this system for a set of initial conditions that could significantly contribute to the maximum eccentricity of planet b without destabilizing the system. We constrain our explored parameter space to the range of masses, semi-major axes, and eccentricities that could reasonably produce this effect.

To guide the construction of our parameter space, we use two approximate expressions based on the two requirements that guide our analysis: stability and secular perturbing strength. For stability, we require that any third planet or binary companion in the system must not disturb the orbits of the two known planets over the lifetime of the system via close encounters, collisions, or ejections. We also require that the secular interactions between planet b and a theoretical object d are strong enough to periodically boost the eccentricity of planet b such that it enters the tidal circularization regime. The two expressions we use have opposite demands and thus help to constrain our parameter bounds: stability requires a less massive and more distant object while secular perturbing strength requires a more massive and closer in object.

The first expression we use (Equation \ref{eq:petro}) is an analytic approximation of stability, developed for two-planet systems in \citet{Pet15}. Systems in which  the inequality is satisfied are expected to evolve secularly without close encounters. Since our system hosts three satellites, this criterion is not directly applicable; however, if the inequality is not satisfied, we can expect the system to be unstable. If objects c and d would not be stable in a two-planet system, adding in planet b is unlikely to improve the stability.

\begin{equation}
\label{eq:petro}
\frac{a_{d}(1-e_{d})}{a_{c}(1+e_{c})}>2.4[\max(\mu_{c},\mu_{d})]^{1/3}\left(\frac{a_{d}}{a_{c}}\right)^{1/2}+1.5
\end{equation}

We can rewrite this criterion by plugging in the known values for Kepler-419c as
\begin{flalign}
\label{eq:pettre}
a_{d}(1-e_{d})& > \nonumber \\   \left[1.7\left(\frac{\max(\mu_d,0.007)}{0.01}\right)^{1/3}\left(\frac{a_{d}}{10 \rm{AU} }\right)^{1/2}+2.1\right] {\rm AU} & \nonumber \\
\end{flalign}

\noindent where $\mu_{c}=m_{c}/m_{\star}$ and $\mu_{d}=m_{d}/m_{\star}$. Parameter sets that satisfy the inequality in Equation \ref{eq:pettre} are considered favorable for stability. The mass, semi-major axis, and eccentricity are limited by this constraint.

The second expression we use (Equation \ref{eq:pericenter}) is an approximation of the angular frequency of precession of a test particle induced by a body with some mass, assuming low eccentricities and low inclinations \citep{Mur99}. We use this expression to roughly approximate the secular effect of object d on planet b, despite our system breaking the test particle and low eccentricity assumptions and, in some iterations, the low inclination assumption. We can compare this to the empirical precession rate of planet b due to planet c measured from an N-body simulation of these two known planets (Figure \ref{fig:two}). In order for the secular effect of object d to be competitive, we require that the frequency of planet b's pericenter precession induced by object d must be greater than 1/10 that by planet c. The expression for this requirement is:

\begin{figure}
\center{\includegraphics{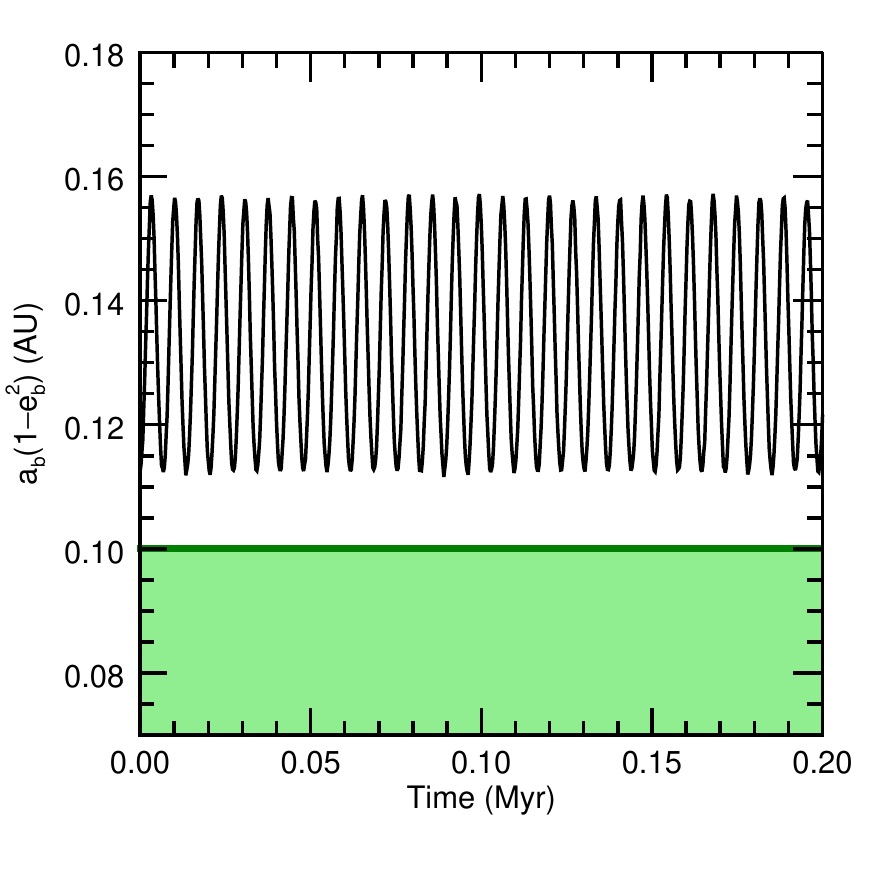}}
\caption{In the two-planet solution to the Kepler-419 system, planet b never reaches a high enough eccentricity to tidally migrate. The black line shows the secular eccentricity cycle of Kepler-419b. In order for significant high-eccentricity tidal migration to occur, the planet must pass into the green region periodically (Equation \ref{eq:criterion}).}
\label{fig:two}
\end{figure}

\begin{equation}
\label{eq:pericenter}
A_{bd}=\frac{1}{4}n_b\frac{m_d}{m_{\star}}\alpha\bar{\alpha} b^{(1)}_{3/2}(\alpha) \geqslant \frac{1}{10}A_{bc},
\end{equation}

\noindent where $A_{bd}$ is the angular frequency of precession of planet b due to planet d, $A_{bc}$ is the angular frequency of precession of planet b due to planet c, $n_b$ is the mean motion of planet b, $m_{\star}$ is the stellar mass, $\alpha=a_b/a_d$, $\bar{\alpha}=\alpha$ (for an exterior perturber), and $b^{(1)}_{3/2}(\alpha)$ is a Laplace coefficient of order unity,

\begin{equation}
\label{eq:laplace}
b^{(1)}_{3/2}(\alpha)=\frac{1}{\pi}\int^{2\pi}_0\frac{\cos\psi d\psi}{(1-2\alpha \cos\psi+\alpha^2)^{\frac{3}{2}}}.
\end{equation}

Substituting $A_{bc}=1.34$ kyr$^{-1}$ measured from the two-planet simulation (Fig. \ref{fig:two}), as well as the mass and semi-major axis of planet b, into Equation \ref{eq:pericenter} yields
\begin{equation}
\label{eq:perisim}
\left(\frac{\mu_d}{0.01}\right)\left(\frac{a_d}{10 \rm{AU} }\right)^{-2} \geqslant \left(\frac{b^{(1)}_{3/2}(\alpha)}{1.19}\right)^{-1}.
\end{equation}

Equation \ref{eq:perisim} favors larger masses and smaller semi-major axes. Equation \ref{eq:pettre}, however, favors larger semi-major axes and lower masses. Intermediate masses and semi-major axes are needed to satisfy both constraints. The range of properties for object d we include after considering these two constraints is provided in Table \ref{tab:init}. Most notably, they are Jupiter,brown dwarf, and late M dwarf-mass objects with semi-major axes less than 10 AU and eccentricities less than 0.45. Due to their approximate nature, we do not use either of these expressions as hard constraints on the parameter space we explore; however, we show in Section \ref{results} that they are useful guidelines because object d parameters that do not satisfy Equation \ref{eq:pettre} tend to lead to unstable systems and object d parameters that do not satisfy Equation \ref{eq:perisim} tend to not excite the eccentricity of planet b high enough.

\begin{deluxetable}{lll}
\tablewidth{3in}
\tablecaption{Range of initial parameters explored for object d \label{tab:init}}
\tablehead{
\colhead{Parameter} & \colhead{Range} & \colhead{Grid spacing}
}
\startdata
$m_d\left(m_{\rm{Jup}}\right)$ & $0.5-115$ & LOG$_{10}$ \\
$a_d\left(\rm{AU}\right)$ & $1.0-10.0$ & Linear \\
$e_d$ & $0-0.45$ & Linear \\
$i_{\rm{mut}}\left(^{\circ}\right)$ & $0-180$ & Linear \\
$\omega_d\left(^{\circ}\right)$ & $0-360$ & Linear \\
$\Omega_d\left(^{\circ}\right)$ & $0-360$ & Linear \\
\enddata
\end{deluxetable}

Although approximate analytic expressions for 4-body secular evolution, such as Laplace-Lagrange excitation, exist, no existing approximation is appropriate here due to the combination of large eccentricities and small semi-major axis ratios. We could use these expressions as rough estimates of the long-term evolution of the system, but they are not reliable enough to determine which systems would be successful and which would not. Thus, we must numerically simulate the system to determine long-term stability and quantify the effect of object d on the eccentricity of planet b.

We run 3495 N-body simulations using the Mercury6 Bulirsch-Stoer integrator \citep{Cha99} with the range of initial conditions for object d defined in Table \ref{tab:init}. We search for stable systems with secular eccentricity oscillations in planet b that periodically result in a large enough eccentricity to tidally migrate.

\section{Results} \label{results}

As discussed in the previous section, in order for a potential Kepler-419d to have contributed to the tidal migration of planet b, two conditions must be satisfied: (1) the amplitude of secular eccentricity oscillations in planet b induced by object d must be strong enough to significantly add to those induced by planet c so that it can get to a high enough eccentricity,
and (2) the system must remain stable at least over the stellar lifetime. These two constraints compete with each other in that (1) favors a high mass, low semi-major axis object, while (2) favors a low mass planet widely separated from the other two planets in the system. If there is an area of parameter space that satisfies both constraints, we expect it to be in an intermediate mass and semi-major axis space where these constraints overlap.

We assess the results of our simulations by considering the satisfaction of these two conditions for each of the parameter sets for object d we test. In the following two subsections, we provide explanations of how we determine whether these conditions are satisfied.

\subsection{Secular Excitation of Planet b} \label{secular}

Following \citet{Soc12a} and \citet{Don14}, we require that planet b must satisfy the inequality in Equation \ref{eq:criterion} in order for tidal migration to have occurred in this system. This inequality expresses the maximum orbital angular momentum that can result in a hot Jupiter ($e\approx0$, $a\leqslant0.1$ AU).

\begin{equation}
\label{eq:criterion}
    \min[a_b(1-e_b^2)]<0.1
\end{equation}

Figure \ref{fig:ecc} shows a tidal circularization track (blacked dashed line) for a planet at the edge of this limit. After migration, a planet following this track would have eccentricity $e=0$ and period $P=10$ days. Observationally, hot Jupiter eccentricities tend to circularize around 0.06 AU \citep{Soc12a}. Since angular momentum is conserved by tidal circularization and tidal dissipation depends strongly on the planet-star separation, the $a(1-e^2)<0.1$ cutoff shown here is a conservative limit \citep{Don14}. In reality, the cutoff for tidal migration on a timescale less than the age of the system could be much stricter; however, because tidal parameters for Jupiter-like planets such as Kepler-419b are not well known, we use the more conservative cutoff of requiring $a_{\rm{final}}=a(1-e^2)$ for planet b to be in the hot Jupiter regime. Object d parameter sets that produce lower  $\min[a_b(1-e_b^2)]$ values are more likely to result in the migration of planet b within the age of the system

\begin{figure}
\center{\includegraphics{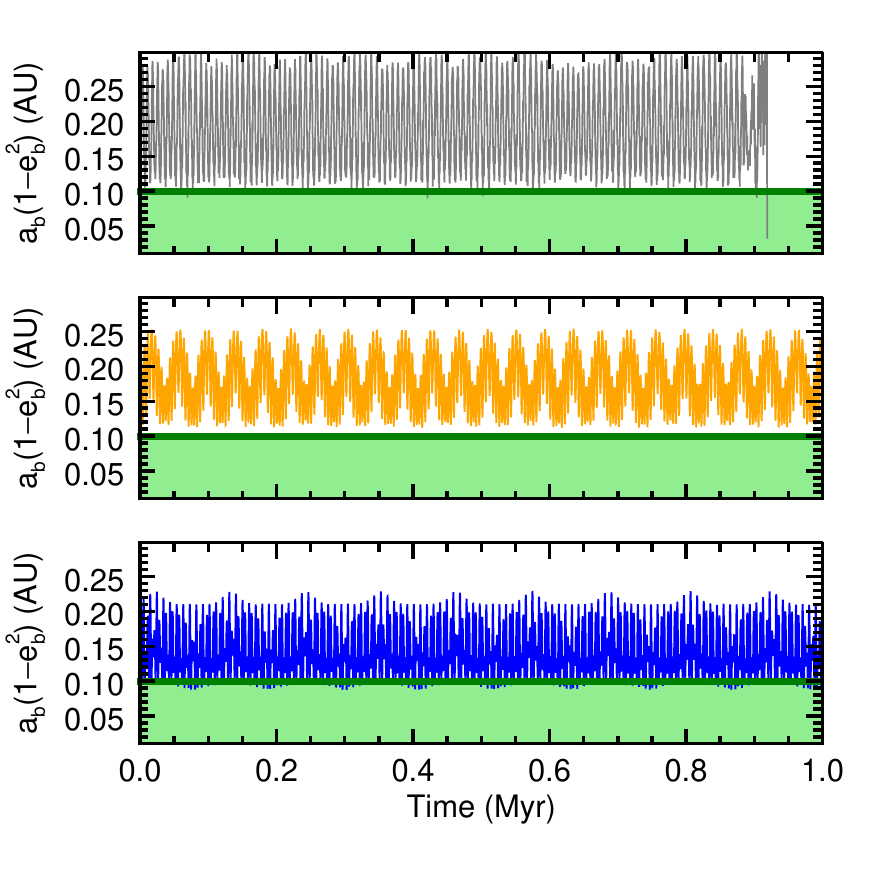}}
\caption{Three representative examples of eccentricity oscillations in Kepler-419b for 1 Myr of simulation for the three planet configuration. Panel 1: unstable system (in this case, a collision between planet b and the central star; see Figure \ref{fig:unstable}). Panels 2 and 3: stable system over the simulation runtime that does not reach the eccentricity cut-off (Equation \ref{eq:criterion}) over many secular cycles.}
\label{fig:three}
\end{figure}

We use 1 Myr simulations to determine which properties of object d allow planet b to reach a sufficient eccentricity to satisfy Equation \ref{eq:criterion}, given that the semi-major axis of planet b remains roughly constant. In order for planet b to undergo tidal migration in its current configuration, it must satisfy Equation \ref{eq:criterion} at some point within its secular cycle. Since, empirically, 1 Myr is enough time to encompass many secular timescales for all of our sets of object d initial conditions, we can rule out any solutions that do not satisfy our criterion in that time. Figure \ref{fig:three} shows a representative example of one simulation that satisfies Equation \ref{eq:criterion} over the first 1 Myr (bottom panel) and one that does not (middle panel). The top panel is discussed in Section \ref{stability}.

For comparison of the effect on the eccentricity of planet b, Figure \ref{fig:two} shows $a_b(1-e_b^2)$, simulated over 10 Gyr with only the known second planet included. Only the first $2*10^5$ years are plotted to show the timescale of the secular cycle. Although the planet's eccentricity is oscillating, it does not satisfy Equation \ref{eq:criterion} in the two-planet solution. Our degree of confidence in that conclusion is very high because the orbital parameters for the two known planets are very precisely constrained (Table \ref{tab:kep}).

\begin{figure*}
\center{\includegraphics{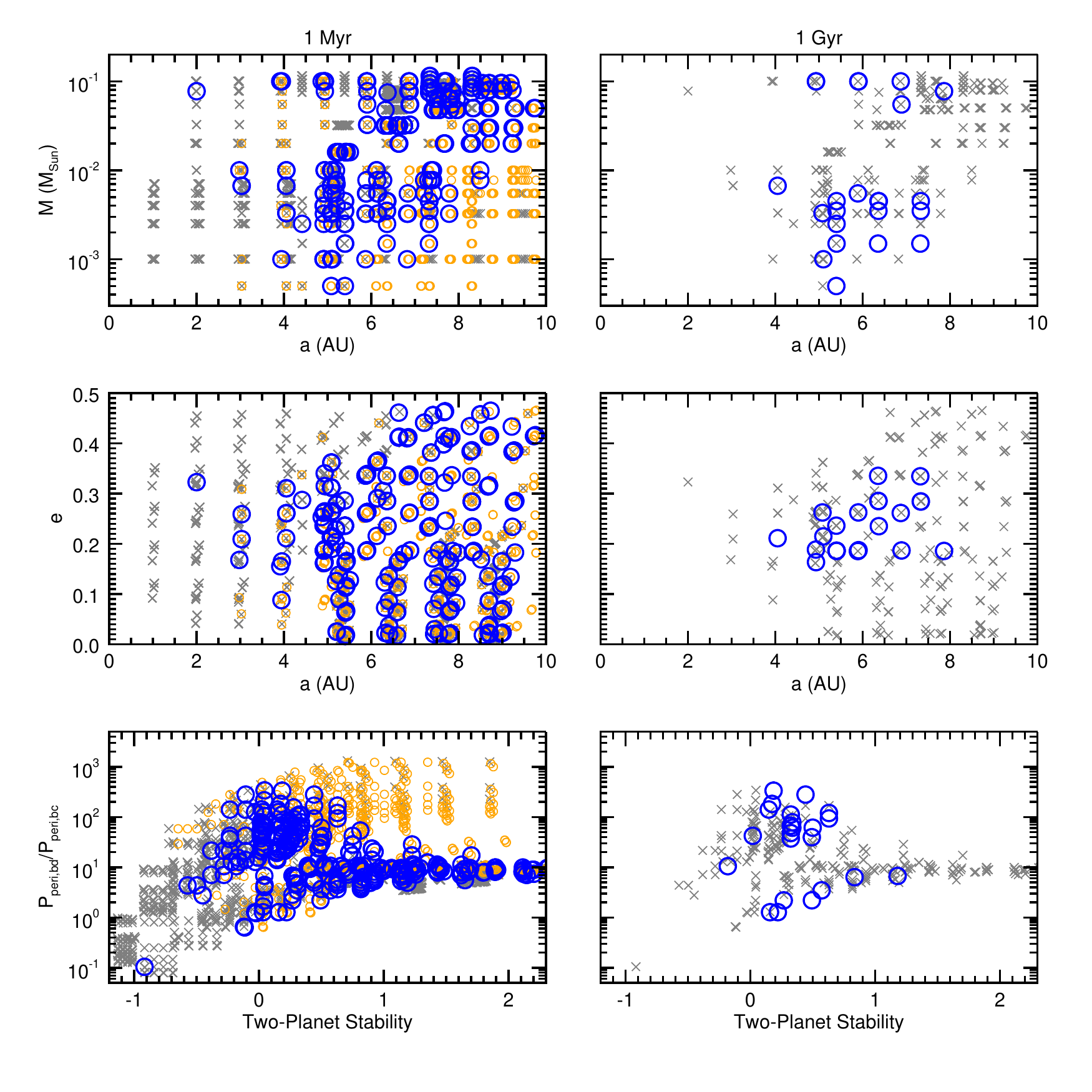}}
\caption{The allowed parameter space for an additional perturbing planet that affects the eccentricity of planet b in the Kepler-419 system is highly limited. The axes of each panel refer to the initial conditions of object d in our simulations at two different integration times: 1 Myr (left) and 1 Gyr (right). Gray crosses represent unstable systems, small orange circles represent stable systems that do not satisfy Equation \ref{eq:criterion}, and large blue circles represent stable systems that do satisfy Equation \ref{eq:criterion}. Solutions that were ruled out after 1 Myr are not included in the 1 Gyr plots. The y-axis on the bottom row of panels is the approximated pericenter precession period of planet b due to object d divided by that due to planet c (Equation \ref{eq:pericenter}). The x-axis is a measure of two-planet stability between planets c and d from \citet{Pet15}, where positive values correspond to stable systems and negative values correspond to unstable systems (Equation \ref{eq:petro}). We can conclude from the right column of panels that a third planet could produce a large enough maximum eccentricity while remaining stable up to 1 Gyr, but only for a small range of parameters. This figure compiles the results from our initial run of 3295 simulations, discussed in Section \ref{sims} as well as our extended run of 200 simulations zooming in on areas of particular interest, discussed in Section \ref{simulations}}
\label{fig:results}
\end{figure*}

The first million years of our simulations partially constrain the allowed parameter space of object d, as demonstrated by the left column of Figure \ref{fig:results}. The orange circles in this figure represent initial parameters of object d that do not sufficiently excite the eccentricity of the inner planet (Equation \ref{eq:criterion} is not satisfied), while blue circles represent successful initial parameter sets. Each panel in Figure \ref{fig:results} shows a projection of the many-dimensional space of object d parameters into two dimensions, which accounts for the overlapping points. We can rule out much of the low mass, large semi-major axis regime with the 1 Myr simulations since that region is dominated by orange circles and devoid of blue circles. Figure \ref{fig:results} also includes the results of 200 follow-up simulations that are discussed in Section \ref{simulations}, which accounts for the varying point density. The initial run of 3295 simulations were roughly uniformly sampled following Table \ref{tab:init}.

\subsection{System Stability} \label{stability}

Because we are seeing Kepler-419b more than a Gyr after its formation (the age of the system is not well constrained beyond that, \citealt{Daw12J}), the system configuration must be stable over the lifetime of the star. Thus, any simulations that go unstable before $\sim$1 Gyr can be ruled out. Unstable simulations are denoted by gray crosses in Figure \ref{fig:results} and an example is shown in the top panel of Figure \ref{fig:three}.

\begin{figure*}
\center{\includegraphics{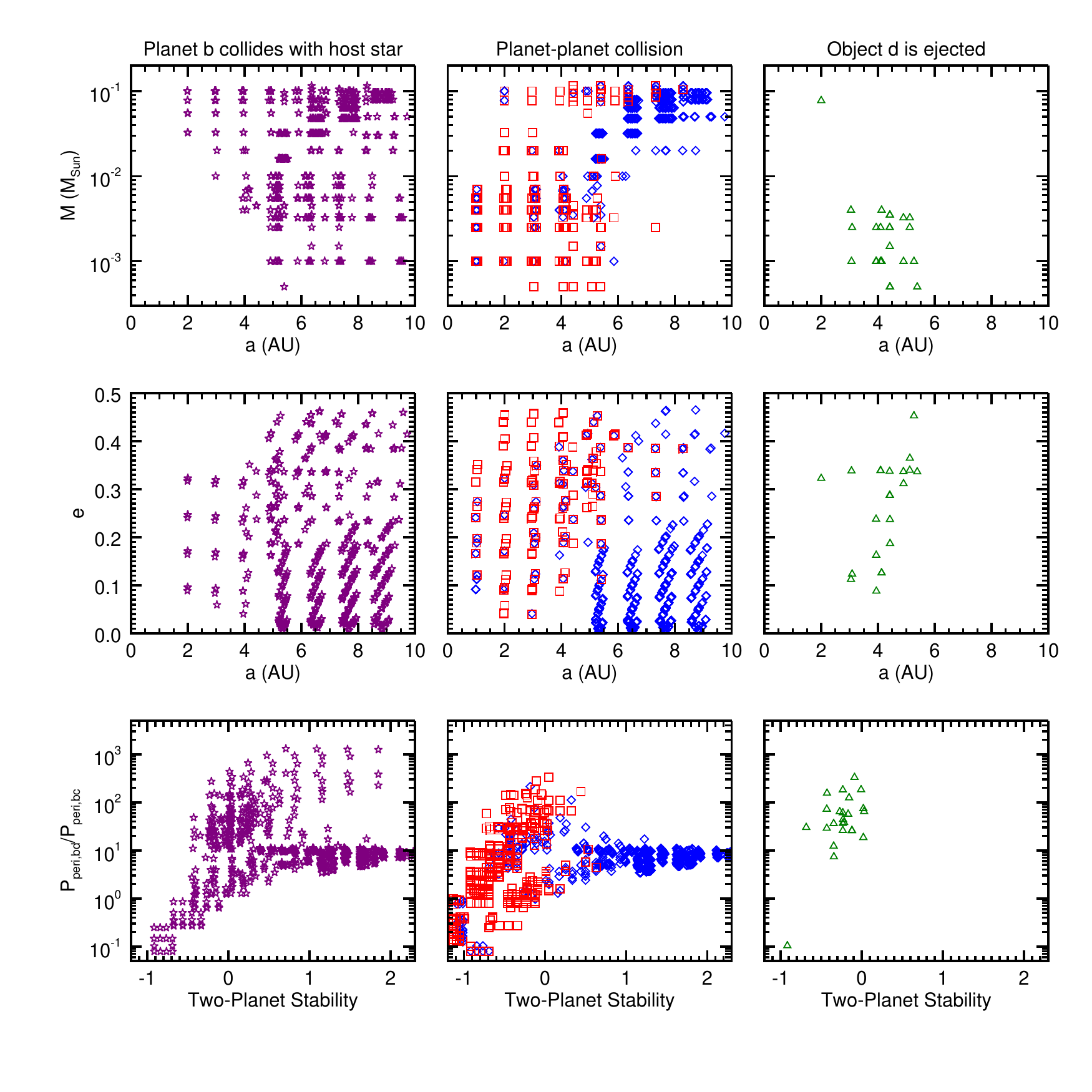}}
\caption{Our simulations go unstable in a variety of ways that are associated with the initial parameters of object d. Left: systems where planet b collides with the central star (purple stars). Middle: Collisions between planets b and c (blue diamonds) and collisions between planet c and object d (red squares). Right: Ejections of object d (green triangles). The parameter space shown here is the same as in Figure \ref{fig:results}.}
\label{fig:unstable}
\end{figure*}

In our simulations, a system configuration is deemed "unstable" if (a) a planet is ejected, (b) a planet collides with the central star, or (c) two planets experience a close encounter, defined as one object moving within a Hill radius of another object. Figure \ref{fig:unstable} provides a visual representation of how each unstable configuration went unstable in the parameter space of object d initial conditions. Green triangles represent systems where object d is ejected, purple stars represent systems where the inner planet collides with the star, and blue diamonds and red squares represent collisions between planets b/c and c/d, respectively. The different ways in which configurations go unstable are somewhat stratified in parameter space, with collisions between planets b and c tending to occur in systems with a high mass, large semi-major axis, small eccentricity object d; collisions between planets c and d tending to occur in systems where object d has a small semi-major axis; and collisions between planet b and the star populating the full parameter space except for the low mass, small semi-major axis object d regime. Planet ejections are rare in our simulations and mostly occur at intermediate semi-major axes and low masses for object d.

\subsection{Simulation Results after 1 Gyr} \label{simulations}

After 1 Myr, the majority of object d parameters could be ruled out because they either went unstable (Section \ref{stability}) or did not sufficiently excite the eccentricity of planet b (Section \ref{secular}). The left column of Figure \ref{fig:results} shows the results of these simulations plotted in the parameter space of object d initial conditions. As stated, gray crosses represent unstable systems, small orange circles represent stable systems that do not satisfy Equation \ref{eq:criterion}, and large blue circles represent systems that remain stable and satisfy Equation \ref{eq:criterion}.

Parameters for object d that did not drive up planet b's eccentricity (orange circles) were not continued beyond 1 Myr. Thus, in the right column of Figure \ref{fig:results} (snapshot at 1 Gyr), the only solutions that are plotted are ones that went unstable after 1 Myr but before 1 Gyr (gray crosses) and ones that satisfy both criteria through 1 Gyr (blue circles). Of the 3295 different sets of parameters for object d we test, 12 satisfy Equation \ref{eq:criterion} while remaining stable for 1 Gyr. These surviving systems are in low mutual inclination orbits with eccentricities between 0.175 and 0.35 and semi-major axes between 4 and 8 AU. Notably, more than half of the successful simulations are in violation of Equation \ref{eq:perisim} (upper cluster of blue circles in the bottom right panel of Figure \ref{results}). This approximation likely fails here due to the broken assumptions discussed in Section \ref{sims} (e.g., massless particle, low eccentricities). While Equation \ref{eq:perisim} was ultimately not used to strictly constrain our parameter space, it did provide a useful baseline when constructing the space.

Of the 12 sets of initial conditions of object d that survive beyond 1 Gyr and satisfy Equation \ref{eq:criterion}, we can define two different populations: one low mass ($5\times10^{-4}-7\times10^{-3}m_{\sun}$) planetary companion group and one high mass ($5.5\times10^{-2}-1.15\times10^{-1}m_{\sun}$) stellar companion group with masses corresponding to brown dwarfs or late M dwarfs. We zoom in on these two areas of parameter space and run 100 simulations in each range with finer resolution in mass, semi-major axis, and eccentricity. These additional 200 simulations have already been included in Figure \ref{fig:results}.

In the planetary-mass group, including both the original parameter sets and the 100 finer-resolution parameter sets, 16 simulations remain stable for 1 Gyr while satisfying Equation \ref{eq:criterion}. In the stellar-mass group, 7 parameter sets are successful through 1 Gyr, bringing the total to 23 successful object d parameter sets. 

Since there are areas of parameter space for a third planet in the Kepler-419 system that contribute strongly to the eccentricity oscillations of planet b without disturbing the stability of the system, we cannot rule out the high-eccentricity tidal migration mechanism for planet b's origin  based on these two criteria. 

\subsection{Chaotic System Evolution} \label{chaos}

\begin{figure}
\center{\includegraphics{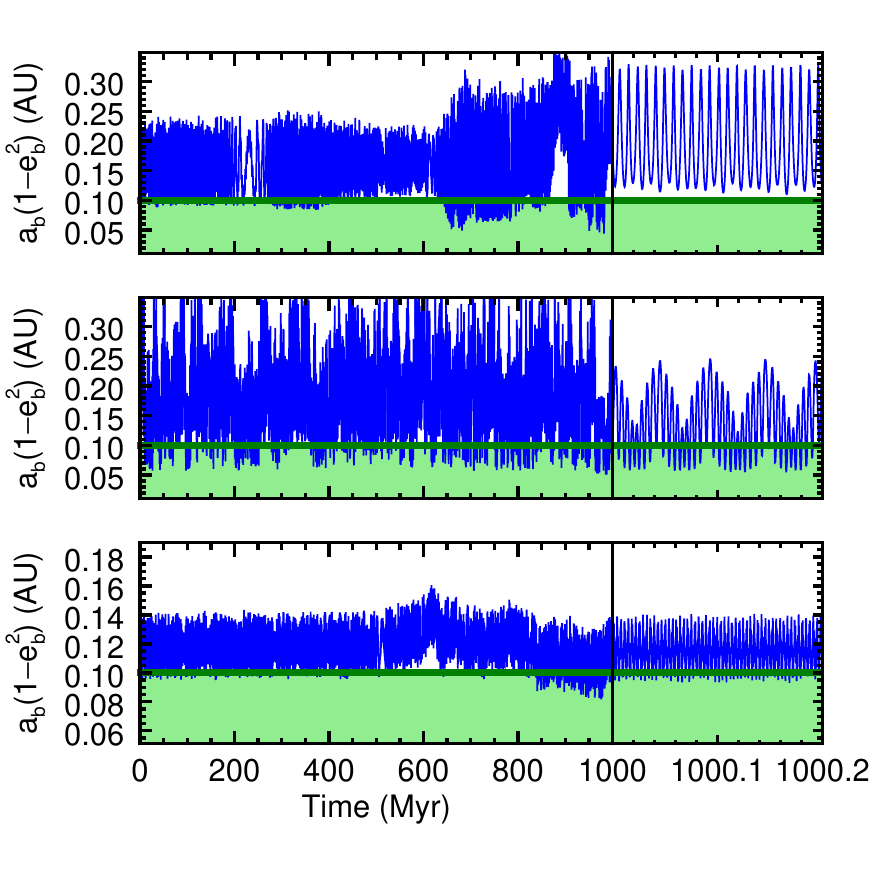}}
\caption{Three representative examples of chaotic evolution in our simulations. These are three of the 23 parameter sets that satisfy Equation \ref{eq:criterion} and survive beyond 1 Gyr. None of these 23 simulations demonstrate smooth secular oscillations throughout their runtime.}
\label{fig:chaos}
\end{figure}

We can divide our stable simulation results into two categories based on the nature of their secular evolution: (1) parameter sets that produce smooth secular oscillations, such as those in the bottom two panels of Figure \ref{fig:three}, and (2) parameter sets that evolve chaotically. Figure \ref{fig:chaos} shows some representative examples of category (2). After 1 Myr of simulation runtime, 47\% of our successful simulations (parameter sets that remain stable and satisfy Equation \ref{eq:criterion}) qualitatively fall into the first category, while the remaining 53\% fall into the second. However, after 1 Gyr, all of the surviving parameter sets are chaotic in their evolution. parameter sets that are not chaotic after 1 Myr either go unstable or are disrupted and become chaotic.

In all successful chaotic simulations, the orbits of planets b and c were not significantly disturbed from their measured values, suggesting the Kepler-419 system could have evolved in a chaotic manner and remain consistent with observations. In fact, some of the chaotically evolving systems reach a much lower minimum $a_b(1-e_b^2)$ than the secularly evolving systems.

Chaotic systems that only passed below $a_b(1-e_b^2)=0.1$ for a small amount of time are treated as systems that do not satisfy Equation \ref{eq:criterion} (orange circles in Figure \ref{fig:results}) because an assumption for that criterion was that planet b would reach its maximum eccentricity periodically throughout the stellar lifetime. Thus we require the $a_b(1-e_b^2)$ to periodically pass below 0.1 for at least half of the simulation runtime. This additional cut is already included in the number of successful simulations we quote in Section \ref{simulations}.

\subsection{Longitude of Pericenter Constraints} \label{peri}

The observed difference in longitude of pericenter $\varpi_c-\varpi_b$ ($\Delta\varpi$) between planet b and planet c is very close to $180^{\circ}$ (see Table \ref{tab:kep}). We are unlikely to have observed the orbits of these two planets in this configuration by chance, which suggests they are locked in a  libration around $\Delta\varpi = 180^{\circ}$ with a small amplitude. By simulating the two-planet case forward in time, we see that this angle does in fact librate about $180^{\circ}$ with an amplitude of $\sim18^{\circ}$ in the best fit two-planet case and $\sim16^{\circ}$ in the coplanar two-planet case. Any additional perturber must avoid disrupting this libration over the stellar lifetime for us to observe it today.

\begin{figure}
\center{\includegraphics{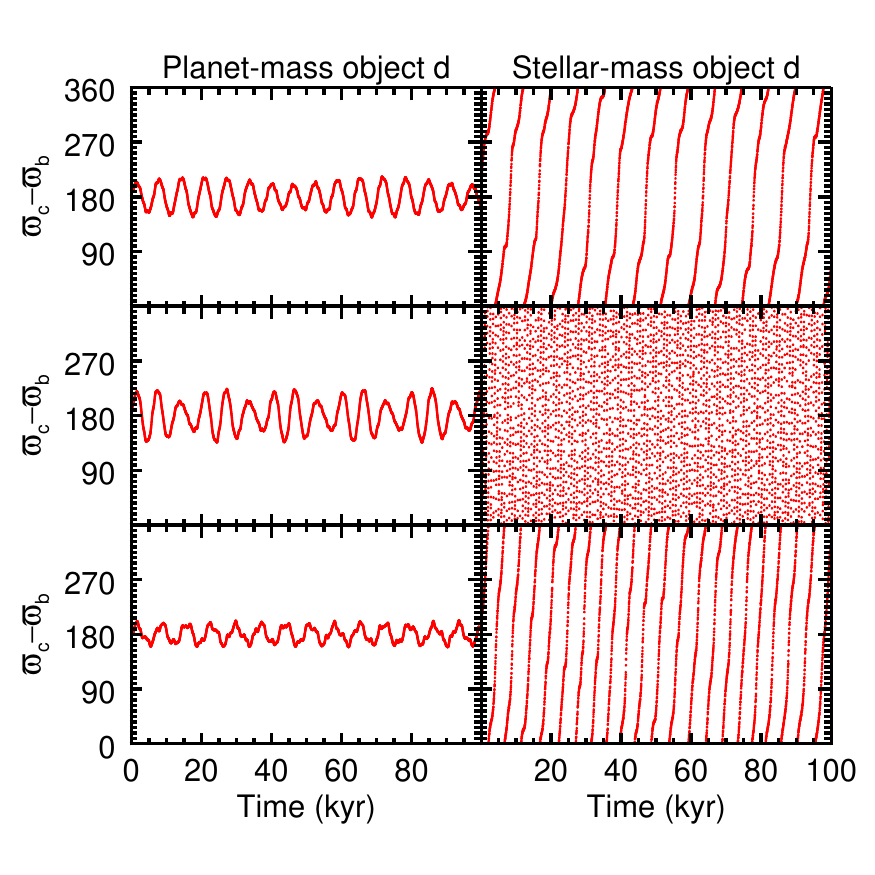}}
\caption{The planetary-mass solutions for a potential object d preserve the libration of $\Delta\varpi$ between planets b and c, while the stellar-mass solutions do not. Only the first 100 kyrs are shown, but 14 of the 16 planetary-mass solutions preserve this libration for the full simulation runtime.}
\label{fig:varpi}
\end{figure}

In Figure \ref{fig:varpi} we show the $\Delta\varpi$ for six representative three-planet configurations that survive longer than 1 Gyr. The systems in the left column have a planetary-mass object d, while the systems in the right column have a stellar-mass object d. Only the planetary-mass perturbers preserve the libration of $\Delta\varpi$ between b and c while satisfying the other conditions necessary for high-eccentricity tidal migration; for all stellar-mass perturbers, the angle circulates. The libration in two of the planetary-mass perturber systems gets disturbed before 1 Gyr, leaving 13 sets of object d parameters that preserve libration of $\Delta\varpi$ between planets b and c over the simulation runtime.

\begin{figure*}
\centering{\includegraphics{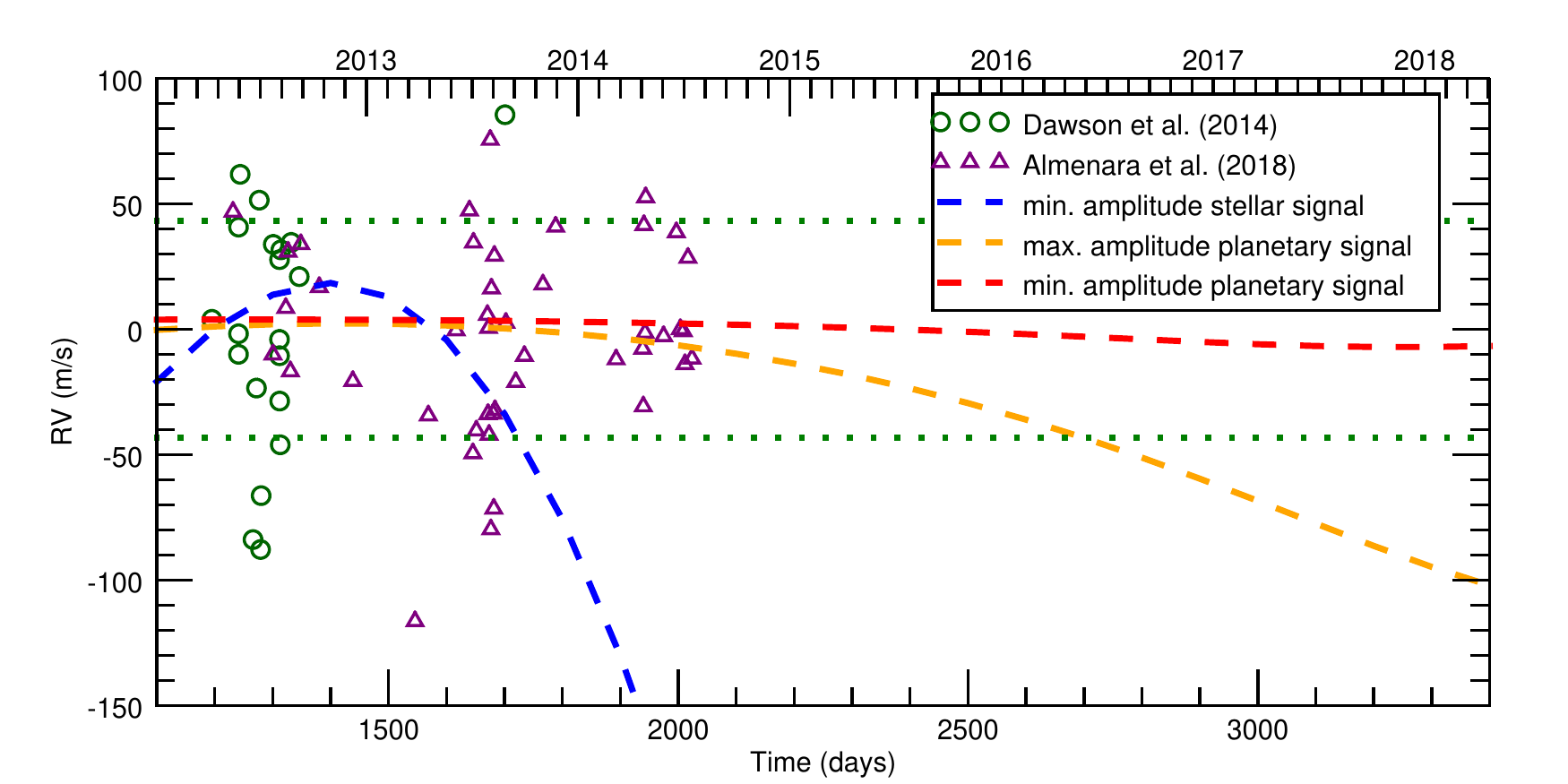}}
\caption{We can rule out all of the stellar-mass group of object d parameters using current RV measurements, and additional measurements up to present day would allow us to rule out some of the planetary-mass group. Green (purple) diamonds (triangles) represent residual RV observations from HIRES (SOPHIE) with the two-planet fit removed \citep{Daw14a,Alm18}. Error bars are not included for clarity, but each point has $\sim$40 ms$^{-1}$ of stellar jitter that dominates the error. RV curves of the minimum-amplitude stellar-mass signal (blue), the maximum-amplitude planetary-mass signal (orange), and the minimum-amplitude planetary-mass signal (red) are extended to present day. Green dotted lines represent the RMS spread in the two-planet fit residuals. The times shown are in BJD-2454833.}
\label{fig:RV}
\end{figure*}

\citet{Alm18} suggest that any additional perturber in the system, would disrupt the libration of $\Delta\varpi$ between the two planets and dismiss four-body secular eccentricity excitation. However, we show that for planetary-mass objects, the $\Delta\varpi$ libration is preserved. Moreover, Since we only observe these planets in a snapshot in time, we cannot be certain that $\Delta\varpi$ is librating in the observed system. Thus, we cannot definitively rule out any areas of parameter space for a potential object d. However, because $\Delta\varpi$ circulates in all of the stellar-mass object d systems and we observe $\Delta\varpi$ very near $180^{\circ}$, we suggest that planetary-mass object d systems may provide a better explanation for the observations.

\section{Observational Constraints} \label{observations}

We can rule out most of the parameter space for object d by applying the two criteria described in Section \ref{results} (Equation \ref{eq:criterion} and stability for 1 Gyr) to our simulations. However, we can also compare the solutions that did satisfy those criteria with current RV (Section \ref{rv}) and TTV (Section \ref{ttv}) measurements of the system to constrain the parameter space even further. As additional observations are published, the allowed parameter space we present here will likely shrink as RV and TTV signals due to a long period planet are detected or ruled out.

\subsection{Radial Velocities} \label{rv}

We model the RV signal of the host star due to object d for each of the 23 sets of initial conditions that satisfy Equation \ref{eq:criterion} and survive for 1 Gyr. In Figure \ref{fig:RV}, we plot a subset of these RV curves as well as the Keck HIRES \citep{Daw14a} and SOPHIE \citep{Alm18} observations of the host star with the best fit two-planet signal subtracted out. The error on these observations includes 40 ms$^{-1}$ of stellar jitter \citep{Daw14a}. The residuals to the two-planet solution are flat, with a root mean square (RMS) spread of $\sim47$ms$^{-1}$. Since the eccentricity perturbations in planet b due to object d are secular, the mean anomaly $M_d$ is not constrained by our simulations. We know the shape and orientation of the orbit, but not the planet's current position along that orbit. Thus, we shift $M_d$ such that the observations fall on the flattest part of the RV curve, minimizing the RV change ($\Delta v$). We plot The minimum-amplitude stellar signal (blue) as well as the minimum-amplitude (red) and maximum-amplitude (orange) planetary signals as dashed lines on Figure \ref{fig:RV}. 

For the planetary-mass group of object d parameter sets, the RV signals due to object d are too small to be detectable with the current data. A small fraction of this group has a large enough RV semi-amplitude, $K\gtrsim47$ms$^{-1}$, to potentially be detected through continued RV monitoring with, for example, the SOPHIE spectrograph. A detailed quantitative analysis of the detectability of these potential planets is beyond the scope of this paper, but Figure \ref{fig:RV} allows us to qualitatively compare the signal to the RMS scatter in the data.

The stellar-mass group produces much larger RV signals. In fact, each of the 7 stellar-mass parameter sets produces a signal with amplitude significantly larger than the RMS scatter of the observations. Thus, a stellar-mass object d that sufficiently excites the eccentricity of planet b cannot be present in the Kepler-419 system.

\subsection{Transit Timing Variations} \label{ttv}

\begin{figure*}
\center{\includegraphics{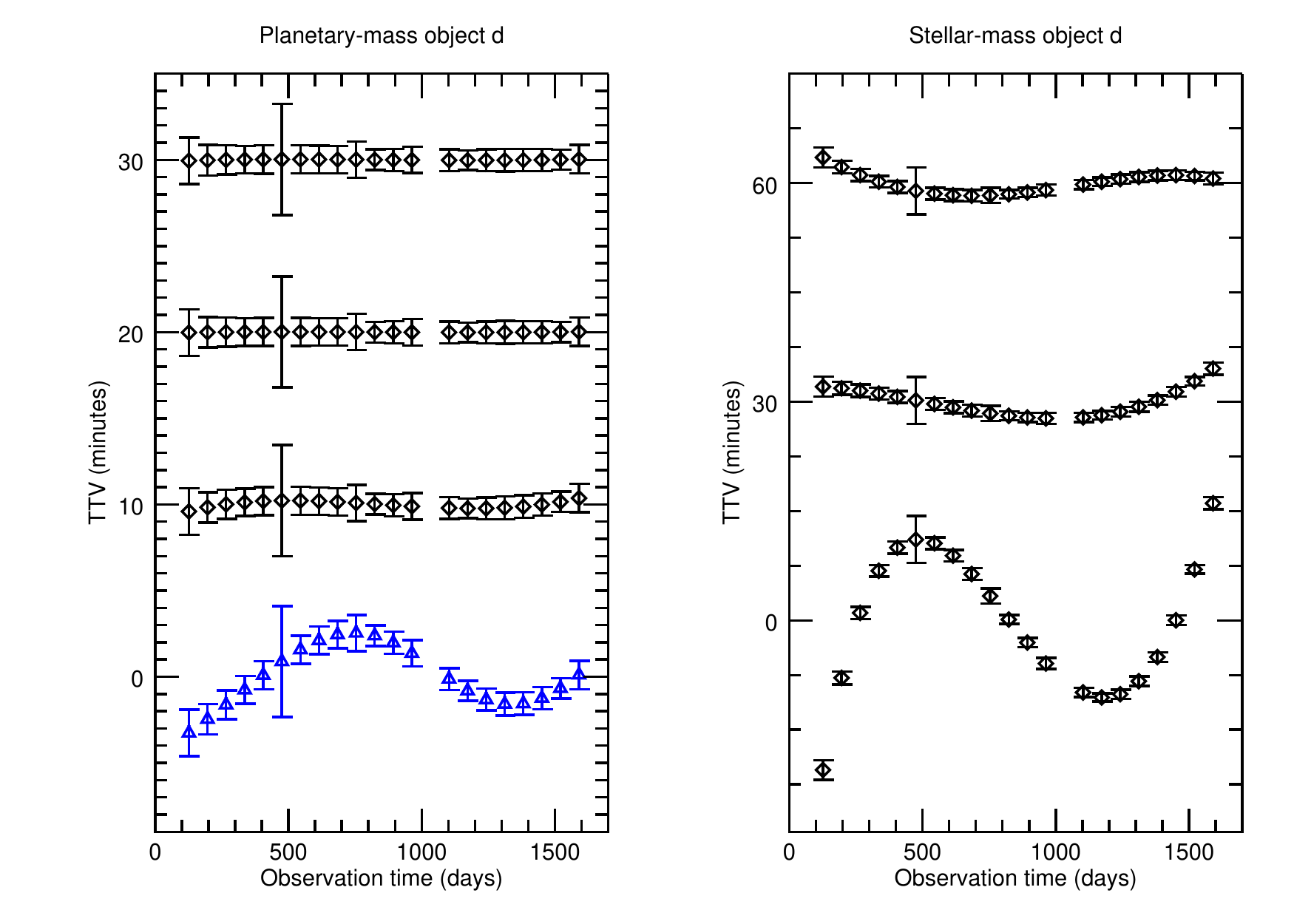}}
\caption{The calculated TTV signals for a potential object d allow us to independently rule out all stellar-mass solutions (right panel), but provide no additional constraints on planetary-mass solutions (left panel). Black diamonds: TTVs of Kepler-419b induced by object d for 6 representative solutions from our simulations at the observed transit times, calculated using the mean anomaly $M_d$ minimizes the TTV amplitude. The error bars represent errors on the time of transit for each observation of the system \citep{Daw14a}. Solutions for which the TTV amplitude is larger than the error bars would have been detected with the current data. Each TTV signal is offset for clarity by 10 minutes for the planetary-mass signals (left) and 30 minutes for the stellar-mass signals (right). Blue triangles: TTV signal due to a planetary-mass object d with an arbitrary mean anomaly $M_d$.}
\label{fig:TTV}
\end{figure*}

We independently model the variation in the time of transit for planet b due to object d for each of the 23 sets of initial conditions that satisfy Equation \ref{eq:criterion} and survive for 1 Gyr. In Figure \ref{fig:TTV}, we plot this TTV signal for several representative examples of object d that encompass the spread of solutions in mass for both planetary and stellar objects. Here we show the TTV signal using the mean anomaly $M_d$ that minimizes TTV amplitude (the same approach used to minimize the RV amplitude in Section \ref{rv}). We qualitatively interpret an amplitude larger than the error bars to be a detectable signal.

In the left column of Figure \ref{fig:TTV}, we show that all of the planetary-mass solutions fail to produce detectable TTVs. In the right column, we show that the TTV amplitude for each of the stellar-mass solutions is larger than the spread of the error bars, independently ruling out stellar-mass objects.

We note that the amplitude of the TTV signal is strongly dependent on the mean anomaly $M_d$. Had we used the $M_d$ that maximized TTV amplitude, nearly all of the planetary-mass signals could be ruled out. To demonstrate this, the blue triangles in Figure \ref{fig:TTV} show one solution with an arbitrary $M_d$. Thus, while we cannot use our TTV analysis alone to rule out any additional solutions, some fine-tuning was required to reach this conclusion. In fact, for the more massive planetary perturbers, about half of the possible values for $M_d$ produce detectable signals, allowing us to rule out configurations with mean anomalies $M_d<150^{\circ}$ and $M_d>310^{\circ}$. For the less massive perturbers ($m_d<\sim 3m_{\rm{Jup}}$), the TTVs would not be detectable for any value of $M_d$

\section{Summary and Discussion} \label{conclusion}

High eccentricity tidal migration has been invoked to explain the origins and properties of hot and warm Jupiters, giant planets orbiting close to their host stars. Although Kepler-419b does not currently have a large enough eccentricity for tidal friction to circularize its orbit, if a perturber were to periodically drive up its eccentricity, it could periodically reach a small enough separation from the star for tidal forces to become important, leading to its current small semi-major axis orbit. Kepler-419c, a coplanar cold Jupiter with a moderate eccentricity, is perturbing the eccentricity of planet b, but not enough to cause high-eccentricity tidal migration on a timescale comparable to the age of the system. However, it is possible for an additional perturbing planet (or small star) to periodically boost the eccentricity of the warm Jupiter into the tidal migration regime.

Using a suite of $\sim3500$ N-body integrations, we explored the parameter space for a potential fourth body in the system that could be exciting the eccentricity of planet b high enough to induce tidal circularization of its orbit without destabilizing the system over the stellar lifetime. We ran these integrations for 1 Gyr and narrowed down the allowed parameter space for this additional perturber to a small region. We also produced RV and TTV curves for all sets of object d parameters that survived for 1 Gyr and compared them to the observations. From the results and analysis of our simulations, we conclude the following:

\begin{enumerate}
    \item Successful sets of initial object d parameters (cases where the system remains stable for 1 Gyr and Kepler-419b reached $a(1-e^2)<0.1$) are constrained to semi-major axes between 4 and 8 AU, eccentricities between 0.175 and 0.35, and mutual inclinations below a few degrees.
    \item The simulated systems with object d initial conditions that survive for 1 Gyr while meeting the eccentricity criterion can be divided into two groups by mass: a planetary-mass group covering $\sim0.5-7$ $m_{\rm{Jup}}$ and a stellar-mass group covering $\sim50-115$ $m_{\rm{Jup}}$.
    \item Many of the simulations that survived to 100 Myrs and all of the simulations that survived to 1 Gyr produced $a_b(1-e_b^2)<0.1$ values that evolved chaotically (not in smooth, consistent secular cycles). The orbits of planets b and c were not significantly disturbed in these cases, so they remain consistent with the observations.
    \item The measured difference in longitude of pericenter between Kepler-419b and Kepler-419c, $\varpi_c-\varpi_b$ ($\Delta\varpi$), is very near 180$^{\circ}$, suggesting the angle is likely librating about this value with a small amplitude \citep{Daw14a}. In the two-planet solution, this angle does librate. All of the stellar-mass group of successful parameter sets for object d failed to preserve this libration for the length of the simulation, along with 2 of the 16 parameter sets from the planetary-mass group. There is a small chance our measurement of $\Delta\varpi$ near 180$^{\circ}$ \citep{Daw14a} is just a coincidence; however, our results suggest that planetary mass perturbers may provide a better explanation for the observations.
    \item The RV and TTV signals for all of the planetary-mass group proved to be consistent with the observations. However, the stellar-mass group could be ruled out because an object in that group would have been detected in both the RV and TTV observations. With the current RV precision and stellar jitter, a subset of the planetary-mass group could potentially be ruled out by extending the RV baseline to the present day if significant linear trends in the RVs are not seen. For the more massive planetary perturbers, TTVs constrain the mean anomaly to the range $150^{\circ}<M_d<310^{\circ}$.
    \item Pending additional RV measurements, we are left with a potential planet between 0.5 and 7 Jupiter masses on a 4-7.5 AU orbit with an eccentricity between 0.18 and 0.35. If a planet in this range of parameter space were to exist in this system, it could produce eccentricity oscillations in Kepler-419b strong enough to periodically pass very close to the host star, allowing tidal friction and, therefore, inward migration.
\end{enumerate}

As additional RV measurements of Kepler-419 are taken, we will be able to further constrain the parameter space allowed for a third planet that could perturb the eccentricity of Kepler-419b enough to be inducing tidal circularization of its orbit. The RV semi-amplitudes of possible solutions for object d are between 5 and 85 ms$^{-1}$. Kepler-419 is an active F-star, making precise RV measurements difficult for this system; however, a longer time baseline combined with detailed stellar noise modeling might put additional constraints on the mass and semi-major axis of any additional hidden planets. Similarly, it may be possible to constrain the allowed parameter space for an additional perturbing planet using future TTV measurements of the system, although they would have to significantly improve upon Kepler's precision. 

Ruling out a four-body tidal migration scenario for Kepler-419b with future observations could lend support to new recent alternative hypotheses for the origin of its eccentric, close-in orbit. \citet{Pet19} posited that Kepler-419b could have acquired its large eccentricity through adiabatic transport during the disk clearing stage. This scenario requires a massive disk and \textit{in situ} formation of the warm Jupiter, but could feasibly explain the observed properties of the system. \citet{Alm18} suggest two alternative scenarios that could increase the eccentricity of planet b: (1) spin-orbit coupling \citep{Cor12} and (2) collision with another planet in a mean-motion resonance. These two scenarios could occur \textit{in situ} or following disk migration. Although we present stable solutions for a four-body system in this paper, planets b and c are very near a stability boundary. Thus, planet b could only have a slightly larger semi-major axis in the past and could not have migrated far while remaining in a stable configuration. \citet{Ant16} claim that most observed Jupiter pairs are dynamically unstable if the inner planets are placed on orbits with semi-major axes larger than 1 AU. This result could suggest that Kepler-419b was and other warm Jupiters with companions were formed \textit{in situ}.

\acknowledgments
We gratefully acknowledge support from grant NNX16AB50G awarded by the NASA Exoplanets Research Program and the Alfred P. Sloan Foundation's Sloan Research Fellowship. This research has also made use of the Exoplanet Orbit Database and the Exoplanet Data Explorer at exoplanets.org. The Center for Exoplanets and Habitable Worlds is supported by the Pennsylvania State University, the Eberly College of Science, and the Pennsylvania Space Grant Consortium. We thank Andrew Shannon for helpful discussions and Cristobal Petrovich, Eric Ford, Jason Wright, Caryl Gronwall, and Jim Kasting for helpful comments. We also thank the anonymous referee for their helpful suggestions.

\bibliography{astrobib}

\end{document}